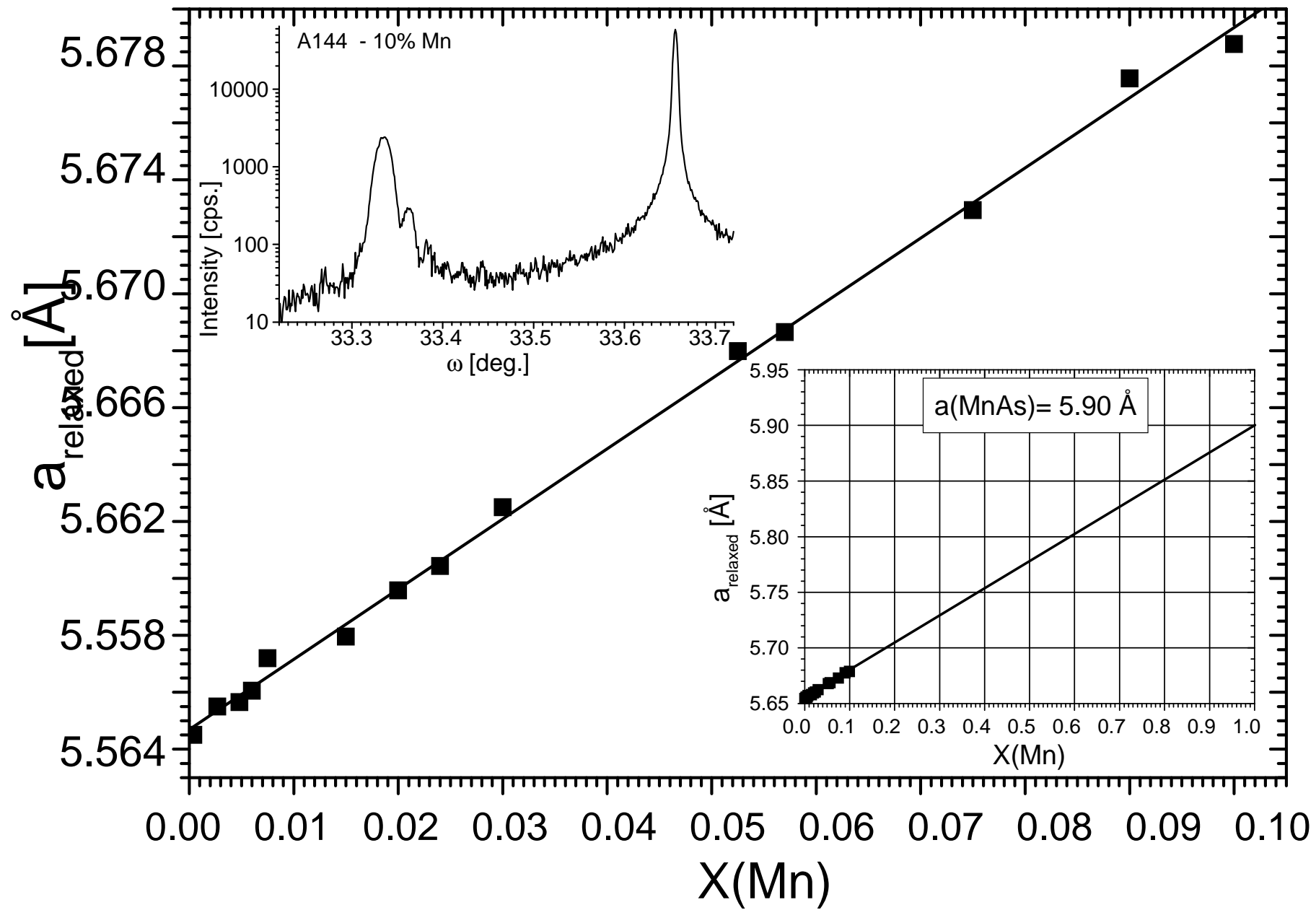

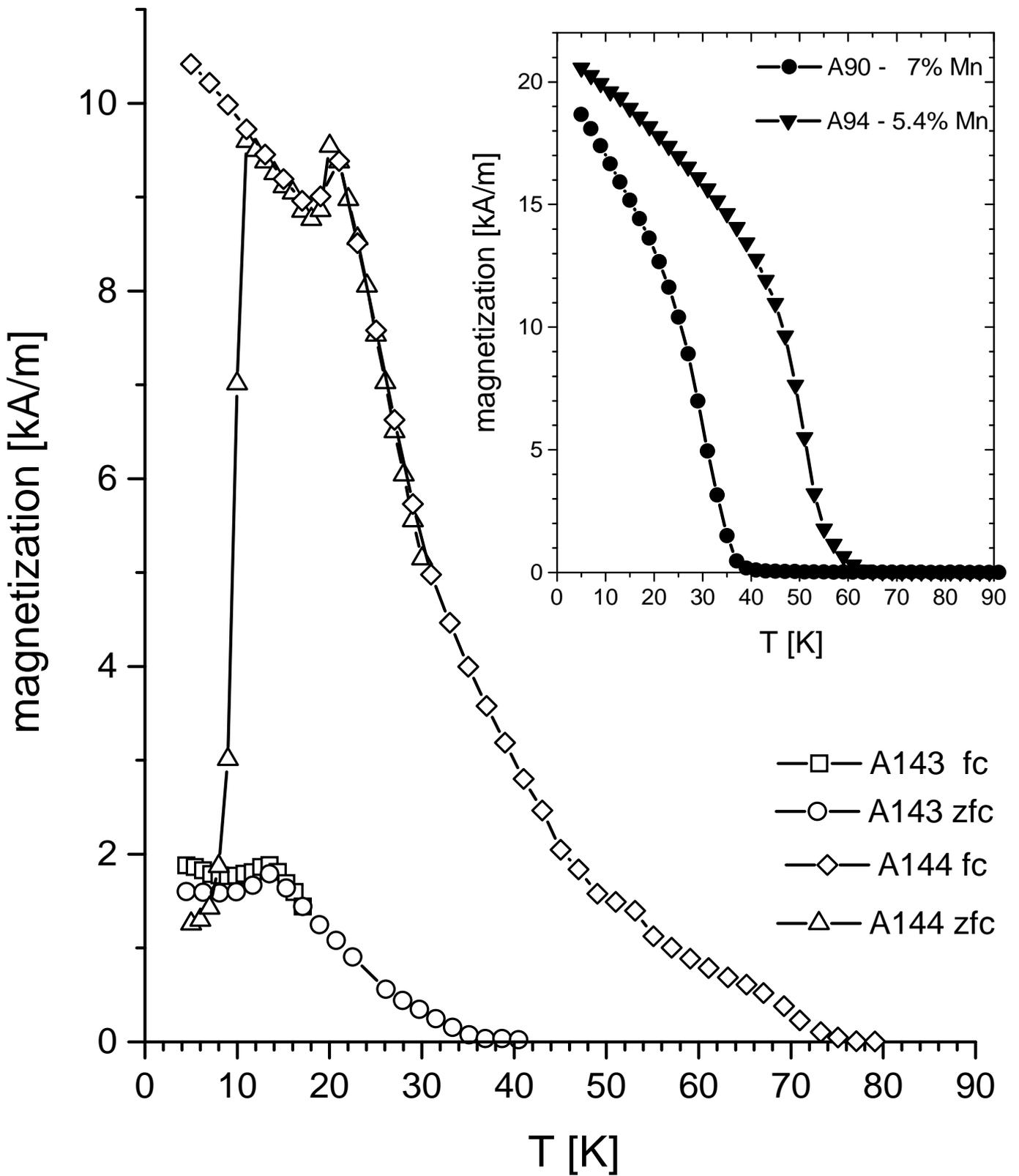

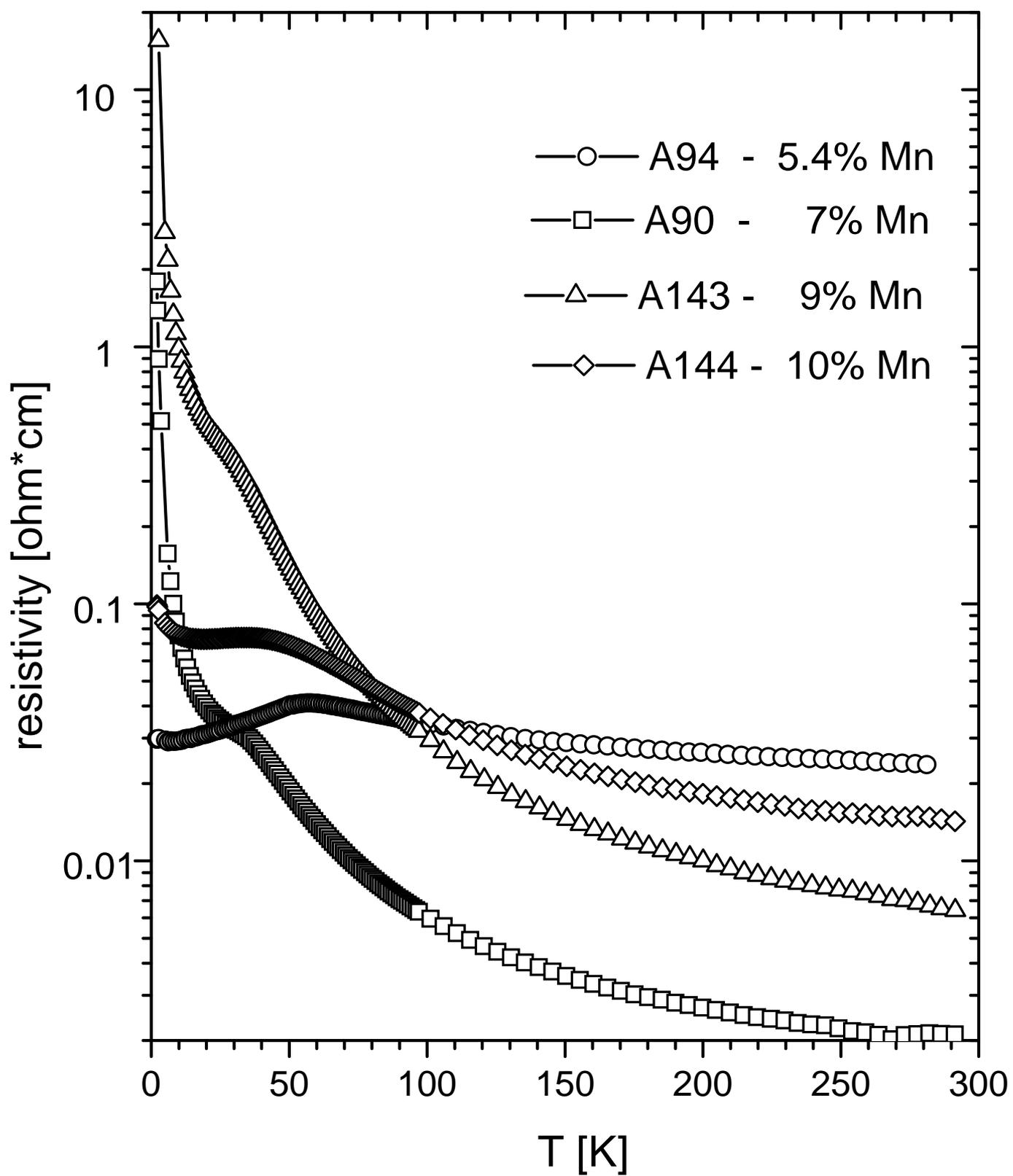

# Structural and magnetic properties of GaMnAs layers with high Mn content grown by Migration Enhanced Epitaxy on GaAs(100) substrates.

J. Sadowski

MAX–lab, Lund University, Box 118, SE–221 00 Lund, Sweden,
phone: (46) 46 222 43 31, fax: (46) 46 222 47 10, e–mail: sadow@maxlab.lu.se
and
Institute of Physics Polish Academy of Sciences, al. Lotnikòw 32/46,
PL–02–668 Warszawa, Poland

R. Mathieu, P. Svedlindh

Uppsala University, Department of Materials Science, SE–751 21 Uppsala, Sweden

J. Z. Domagala, J. Bak – Misiuk, K. Swiatek

Institute of Physics Polish Academy of Sciences, PL–02–668 Warszawa, Poland

M. Karlsteen, J. Kanski, L. Ilver, H. Åsklund, U. Södervall

Department of Experimental Physics, Chalmers University of Technology
and Göteborg University, SE–412 96 Göteborg, Sweden

We have grown the ferromagnetic semiconductor GaMnAs containing up to 10% Mn by migration enhanced epitaxy at a substrate temperature of 150ºC. The alternate supply of $As_2$ molecules and Ga and Mn atoms made it possible to grow single crystalline GaMnAs layers at very low substrate temperature, at which conventional molecular beam epitaxial growth under excess As supply is not possible due to As condensation. Secondary ion mass spectroscopy and X–ray diffraction measurements confirmed a higher Mn content in the films grown by this method in comparison to the GaMnAs layers grown by low temperature molecular beam epitaxy. The lattice constant of hypothetical zinc–blende structure MnAs is determined to be 5.9Å, which deviates somewhat from previously reported values. This deviation is ascribed to growth–condition dependent density of point defects. It is stressed that this effect must be taken into account for any assessment of Mn content from X–ray diffraction data. Magnetization measurements showed an onset of ferromagnetic ordering around 75 K for the GaMnAs layer with 10% Mn. This means that the trend of falling Curie temperatures with increasing Mn concentrations above 5.5% is broken. We tentatively assign this to the variation of the carrier concentration, including contributions from donor and acceptor centers formed by antisite defects and Mn doping, and increased density of magnetically active Mn ions.

GaMnAs is a diluted magnetic semiconductor (DMS) having the highest paramagnetic–to–ferromagnetic transition temperature (Curie temperature – $T_c$) within the family of DMS materials known so far. Ohno et al.[1] reported a record $T_c$ of 110 K for GaMnAs containing 5.3% Mn. Typically, $T_c$ reaches maximum for Mn concentrations close to this value and for higher Mn content $T_c$ decreases. The reason for this non–monotonic variation of $T_c$ with Mn concentration is still being discussed[2–5]. It is well established that the ferromagnetism in GaMnAs is caused by magnetic exchange interaction between charge carriers (holes) and Mn magnetic moments[1,3] and the $T_c$ value strongly depends on both the Mn content and concentration of holes[3]. A high density of holes in GaMnAs (up to $10^{21}$cm$^{-3}$) is supplied by the Mn acceptor centers. Calculations of Dietl et al.[3,4] based on the Zener model in the mean field approximation predict a further increase of $T_c$ (up to 300 K) in GaMnAs with 10% Mn, assuming that this high Mn content can be combined with a correspondingly high concentration of holes. Other models applied recently to GaMnAs, like the density functional theory in the local spin density approximation[5] stress the importance of the disorder connected with the nonuniform distribution of both As antisite defects and Mn ions in GaMnAs.

In this work we report on a growth technique which makes it possible to obtain GaMnAs films containing more than 8% Mn. In low temperature molecular beam epitaxial (LT MBE) growth at substrate temperatures ($T_s$) between 170º C and 250º C we, like other groups[1,6], always observe MnAs clustering for Mn fluxes corresponding to more than 8% Mn content in GaMnAs. Below 8% Mn and substrate temperature not higher than maximum growth temperature no evidence for Mn segregation was found. However as Mn is randomly distributed in Ga sites of the GaAs lattice the local variations of Mn content are possible. As shown recently by theoretical models[5,7] both composition and hole density fluctuations can lead to increased ferromagnetic transition temperature.

It is well known that an increase of Mn content in GaMnAs must be accompanied by a decreased of the growth temperature[6,8,9]. In the case of MBE growth using an As$_2$ valved cracker source and low As$_2$ overpressures, successful growth at temperatures down to about 170 ºC has been reported[9]. Below this substrate temperature conventional LT MBE growth is not possible due to condensation of excess arsenic on the growing film surface. To be able to decrease the substrate temperature further, without having excess arsenic adsorbed, the arrival rate of As$_2$ must be controlled precisely. We achieved this by alternate shuttering of the As$_2$, Ga and Mn sources, and were thus able to decrease the growth temperature down to 150 ºC. This modification of the MBE growth of III–V compounds is known as migration enhanced epitaxy (MEE) [10].

During MEE growth experiments performed at $T_s$=150ºC, with properly adjusted fluxes of the constituent elements, no evidence was found in RHEED for any formation of MnAs inclusions at Mn fluxes, at which conventional LT MBE growth mode always resulted in formation of MnAs clusters.

After mounting into the MBE system the epi–ready GaAs(100) substrates were subjected to the typical growth procedure of high temperature GaAs buffer. The substrate temperature was then ramped down to 150º C during about 2 hours. When the substrate temperature was lower than 400º C the As shutter and the cracker valve were closed. The As source temperature was then decreased to the value corresponding to an As$_2$ flux of 0.33 As molecular layer per second. The substrate temperature during growth of GaMnAs was 150º C, as measured by a thermocouple in contact with the molybdenum holder. This temperature was also calibrated by observing in RHEED the onset of amorphous arsenic deposition during slow cooling of the GaAs substrate in an As$_2$ flux. The Ga + Mn flux ratios were set to a value corresponding to growth of 1 ML GaMnAs per 4 seconds. This was calibrated by means of RHEED oscillations[9,11] (see below) on a test sample during LT MBE growth at a substrate temperature of 200º C. RHEED patterns observed during the MEE growth revealed a two– dimensional surface morphology with a (1x2) reconstruction typical for a GaMnAs surface[6,12].

The structural quality of the MEE grown GaMnAs films is very good, as can be seen in the XRD results (upper inset in Fig. 1). Reciprocal space maps were measured for all samples listed in Table1, indicating that in each case the GaMnAs layer is coherently strained to the GaAs substrate. The clear X–ray interference fringes prove a high structural perfection of the GaMnAs layers. The film thickness calculated from the angular spacing of the fringes corresponds well to the growth rate of precisely one monolayer per one MEE growth cycle. Both strained and relaxed lattice parameter values of GaMnAs films are listed in Table 1. Relaxed (bulk) lattice constant values were calculated assuming the GaMnAs elasticity constants $C_{11}$ and $C_{12}$ to be the same as for GaAs. The Mn content in the MEE GaMnAs layers was estimated by four methods (see Table 1):

(i) Mn flux calibrations based on growth rate differences between GaAs and GaMnAs as measured by RHEED oscillations,
(ii) SIMS measurements,
(iii) Auger microprobe measurements
(iv) lattice constant measurements

The in−situ method (i) can be applied only in the case of MBE grown GaMnAs. It provides values of GaMnAs composition with an accuracy better than 0.5%, as the damping of RHEED oscillations observed both for LT GaAs and subsequently growth GaMnAs is very small[9]. Since the MBE growth proceeds at low temperature (below 250° C) the sticking coefficients of Ga and Mn are equal to one. Due to the As excess, all the Mn atoms are incorporated into the growing film and the increase of the growth rate is proportional to the number of Mn atoms supplied. Thus, assuming that all Mn atoms are incorporated into the Ga sites, the Mn content in the GaMnAs ternary compound is obtained.

Methods (ii) and (iii) need initial calibrations, i.e. they can give quantitative results assuming that for at least one sample the Mn content is known and set as the reference for all the other measurements.

Method (iv) − lattice constant measurements is widely used by different groups working on GaMnAs[13, 14]. Obviously results based on this method depend strongly on a correct value of both: the lattice parameter of a hypothetical zinc−blende MnAs and LT GaAs as well as on the precision of XRD measurements. It was shown by Shimizu et al.[15] (and confirmed in our XRD results), that the GaMnAs lattice constant depends not only on Mn content but also on the growth conditions, such as As overpressure and substrate temperature. This effect is also well known in the case of LT GaAs[16, 17]. Due to the different concentration of excess As defects[18] like As interstitials, Ga vacancies and other defects, typical for LT GaAs, the lattice constant of GaMnAs with the same Mn content can by slightly different depending on growth temperature and As overpressure[15]. Thus estimating the GaMnAs composition from the lattice constant measurements without substantial errors can be done only in the case of LT MBE growth at exactly the same conditions concerning substrate temperature and As overpressure. The results of lattice constant measurements for MBE grown GaMnAs layers grown at 200° C and $As_2$/Ga flux ratio equal to 4.5, with compositions from 0.04% Mn to 7% Mn measured by SIMS and at compositions exceeding 1% Mn also by RHEED intensity oscillations are shown in Fig. 2.

The lattice constant of GaMnAs in this composition range follows Vegard's law:

(1)  $a(Ga_{1-x}Mn_xAs) = 5.65469 + 0.24661*X$   [Å]

this gives the lattice parameter of low temperature GaAs ($a_{LTGaAs}$) equal to 5.65469 Å and that of zinc−blende MnAs ($a_{MnAs}$) equal to 5.90 Å.

In most previous papers on GaMnAs $a_{MnAs}$ is assumed to be 5.98 Å[1,12,14], though smaller values (close to 5.88 Å) have also been used[19]. Also theoretical data show quite varying values: using density functional calculations within the local spin density approximation, Sanvito et. al.[20] obtained a value between 5.6 Å and 5.7Å, while Shirai et al[21], using another method (full−potential linearized augmented−plane−wave) obtained 5.9 Å. This latter value is consistent with our extrapolated zinc−blende MnAs lattice constant.

Magnetisation measurements were performed in a Quantum Design MPMS5 Super–conducting Quantum Interference Device (SQUID) magnetometer. Fig. 2 shows the low field (H = 20 Oe) zero–field cooled (ZFC) and field cooled (FC) magnetizations vs. temperature for the $x = 0.09$ and $x = 0.10$ films. For reference we also show the results for $x = 0.055$ and $x = 0.07$. The two latter are typical for the $Ga_{1-x}Mn_xAs$ system, showing ferromagnetic transition temperatures decreasing with increasing Mn content ($T_c$ = 55 K and 36 K, respectively). For Mn concentrations larger than $x = 0.07$, the magnetization curves show broader features, without sharp transitions to a ferromagnetically ordered state. The saturation magnetization values are also lower for the MEE films in comparison to MBE grown films. The results for $x = 0.09$ seem to follow the trend that the magnetic properties become weaker for concentrations larger than $x = 0.055$. On the other hand, the results for $x = 0.10$ are quite encouraging, as the onset of a magnetically ordered state occurs at a higher temperature, near 75 K. The broad magnetization features seen in Fig. 2 are tentatively attributed to the sample inhomogeneity, related to the Mn and/or hole concentration. In spite of the higher Mn content, the concentration of holes in MEE GaMnAs samples may be lower than in LT MBE samples due to the lower growth temperature and in consequence higher compensation of Mn acceptors. In addition, MEE samples having similar Mn content (9% and 10% Mn) exhibit clear differences with respect to their magnetic and electric properties. This indicates significant differences in hole concentration between the samples. It might also explain the scatter in $T_c$ reported in the literature for samples having the same nominal Mn concentration. König et al.[22] recently presented a theory for carrier induced ferromagnetism in diluted magnetic semiconductors according to which also a non–monotonous variation of $T_c$ with the concentration of holes is predicted. It is thus likely that a proper control of defect density and thus of the density of charge carriers may lead to larger values of $T_c$ [3,4].

In the resistivity vs. temperature results shown in Fig. 3 it can be seen that the resistivity of the MBE samples exhibits distinct features at $T_c$; the same features appear for the MEE samples around the magnetic ordering temperature. It is well known that $Ga_{1-x}Mn_xAs$ exhibits metallic behaviour in a confined Mn composition range around $0.02 < x < 0.055$. For other alloy compositions insulator–like temperature dependence is reported[1,15,23]. Our samples with $x = 0.07$ and $x = 0.09$ follow this trend, exhibiting an overall insulator–like behaviour. However, the $x = 0.10$ sample exhibits a metallic–like resistivity at low temperatures, similar to that observed for $x = 0.055$.

In conclusion, we report successful growth of ferromagnetic III–V semiconductor $Ga_{1-x}Mn_xAs$ films by migration enhanced epitaxy at a very low substrate temperature (150°C). At this temperature, conventional LT MBE growth is not possible due to condensation of excess $As_2$. The MEE–grown layers have a Mn content higher than what is possible to attain by MBE growth. While $x = 0.09$ follows the 'trend' of MBE grown layers, with lower ferromagnetic ordering temperature than $x = 0.07$ and an overall insulator– like behaviour with a slightly higher resistivity than for $x = 0.07$, the $x = 0.10$ sample exhibits magnetic ordering at much higher temperature (near 75 K), as well as metallic–like low temperature resistivity characteristics.

This work has been supported by the Swedish Natural Science Research Council. One of the authors (J.S.) would like to thank MAX–Lab, Lund University for a financial support.

Table 1. Parameters of MEE and MBE GaMnAs samples.

| # | Thickness [μm] | growth method | a [Å] | $a_{bulk}$ [Å] | $x_{Mn}$ RHEED [%] | $x_{Mn}$ SIMS [%] | $x_{Mn}$ XRD [%] |
|---|---|---|---|---|---|---|---|
| A94 | 0.8 | MBE | 5.67936 | 5.6670 | 5.4 | – | 5.0 |
| A90 | 0.9 | MBE | 5.68613 | 5.67058 | 7 | – | 6.5 |
| A143 | 0.3 | MEE | 5.69940 | 5.67756 | 9 * | 9.2 | 9.3 |
| A144 | 0.3 | MEE | 5.70170 | 5.67877 | 10 * | – | 9.8 |

* For MEE growth, direct measurements of composition from RHEED oscillations are not possible. In this case the composition was extrapolated from RHEED measurements obtained for MBE grown GaMnAs assuming that the increase of Mn content is proportional to the increase of Mn/Ga flux ratio.

**Figure captions:**

**Fig. 1.** Lattice parameter of GaMnAs vs. composition. Upper inset shows the X–ray diffraction results – (004) Bragg reflection for GaMnAs 10% MEE sample. Lower inset shows the extrapolation of GaMnAs lattice parameter to the hypothetical zinc–blende MnAs.

**Fig. 2.** Magnetization versus temperature curves measured by SQUID.
    fc – field cooled measurements
  zfc – zero–field cooled measurements
Results for GaMnAs samples containing 5.4% Mn and 7% Mn are shown in the inset.

**Fig. 3.** Resistivity vs. temperature curves. The transition temperature from paramagnetic–to–ferromagnetic phase can be recognised as a local maximum.